\documentclass[12pt]{article}
\usepackage{a4}
\usepackage{epsfig}

\def\ee{\mbox{e}^+\mbox{e}^-}

\def\pz{\phantom{0}}
\def\pzz{\phantom{00}}

\def\Z0{\mbox{Z}^0}
\def\LQ{\mbox{LQ}}
\def\X{\mbox{X}}
\def\ETMISS{E_{\rm T}\!\!\!\!\!\!\!/\;\;}
\def\ppbar{\overline{\mbox p}\mbox{p}}

\def\qqbar{\mbox{q}\overline{\mbox{q}}}
\def\fg{f_{\gamma/e}}
\begin{document}
\begin{titlepage}
\begin{center}{\large   EUROPEAN ORGANIZATION FOR NUCLEAR RESEARCH
}\end{center}\bigskip
\begin{flushright}
       CERN-EP-2001-040   \\ 17 May 2001
\end{flushright}
\bigskip\bigskip\bigskip\bigskip\bigskip
\begin{center}{\huge\bf  
\boldmath
Search for Single Leptoquark and Squark Production in
Electron-Photon Scattering at $\sqrt{s_{\rm ee}}=189$~GeV at LEP
\unboldmath
}\end{center}\bigskip\bigskip%
\begin{center}{\LARGE The OPAL Collaboration
}\end{center}\bigskip\bigskip
\bigskip\begin{center}{\large  Abstract}\end{center}
A search for first generation scalar and vector leptoquarks (LQ) 
as well as for squarks ($\tilde{\mbox{q}}$) in R-parity violating SUSY models 
has been performed using $\ee$ collisions collected 
with the OPAL detector at LEP at an e$^+$e$^-$ centre-of-mass energy 
$\sqrt{s_{\rm ee}}$ of 189~GeV.
The data correspond to an integrated luminosity of about 160~pb$^{-1}$.
The dominant process for this search is 
$\mbox{eq}\to\LQ/\tilde{\mbox{q}}\to\mbox{eq},\nu\mbox{q}$,
where a photon, which has been radiated by one of the beam electrons, serves
as a source of quarks.
The numbers of selected events found in the two decay channels are 
in agreement with the expectations
from Standard Model processes. This result allows to set
lower limits at the 95~\% confidence level 
on the mass of first generation scalar and vector leptoquarks, and of
squarks in R-parity violating SUSY models. 
For Yukawa couplings $\lambda$ to fermions larger than 
$\sqrt{4\pi\alpha_{\rm em}}$, the mass limits range from
121~GeV$/c^2$ to $175$~GeV$/c^2$ 
($149$~GeV$/c^2$ to $188$~GeV$/c^2$) depending on the branching ratio $\beta$
of the scalar (vector) leptoquark state.  
Furthermore, limits are set on 
the Yukawa couplings $\lambda$ for leptoquarks and $\lambda'_{1jk}$ for 
squarks, and on $\beta$ as a function of the scalar leptoquark/squark mass.
\bigskip\bigskip\bigskip\bigskip
\bigskip\bigskip

\begin{center}{\large
(To be submitted to Eur. Phys. J. C)
}\end{center}
\end{titlepage}
\begin{center}{\Large        The OPAL Collaboration
}\end{center}\bigskip
\begin{center}{
G.\thinspace Abbiendi$^{  2}$,
C.\thinspace Ainsley$^{  5}$,
P.F.\thinspace {\AA}kesson$^{  3}$,
G.\thinspace Alexander$^{ 22}$,
J.\thinspace Allison$^{ 16}$,
G.\thinspace Anagnostou$^{  1}$,
K.J.\thinspace Anderson$^{  9}$,
S.\thinspace Arcelli$^{ 17}$,
S.\thinspace Asai$^{ 23}$,
D.\thinspace Axen$^{ 27}$
% ...
}\end{center}\bigskip
\bigskip
$^{  1}$School of Physics and Astronomy, University of Birmingham,
Birmingham B15 2TT, UK
\newline
$^{  2}$Dipartimento di Fisica dell' Universit\`a di Bologna and INFN,
I-40126 Bologna, Italy
\newline
$^{  3}$Physikalisches Institut, Universit\"at Bonn,
D-53115 Bonn, Germany
\newline
% ...
$^{  5}$Cavendish Laboratory, Cambridge CB3 0HE, UK
\newline
% ...
$^{  9}$Enrico Fermi Institute and Department of Physics,
University of Chicago, Chicago IL 60637, USA
\newline
% ...
$^{ 16}$Department of Physics, Schuster Laboratory, The University,
Manchester M13 9PL, UK
\newline
$^{ 17}$Department of Physics, University of Maryland,
College Park, MD 20742, USA
\newline
% ...
$^{ 22}$Department of Physics and Astronomy, Tel Aviv University,
Tel Aviv 69978, Israel
\newline
$^{ 23}$International Centre for Elementary Particle Physics and
Department of Physics, University of Tokyo, Tokyo 113-0033, and
Kobe University, Kobe 657-8501, Japan
\newline
% ...
$^{ 27}$University of British Columbia, Department of Physics,
Vancouver BC V6T 1Z1, Canada
\newline
% ...
Tel Aviv 69978, Israel.
\section{Introduction}
Leptoquarks (LQ) are coloured spin 0 or spin 1 particles carrying both
baryon (B) and lepton (L) quantum numbers. They appear in many extensions
of the Standard Model as a consequence of the symmetry between the lepton
and quark sectors. The Buchm\"uller-R\"uckl-Wyler (BRW) model~\cite{bib-buch}
used in this paper assumes lepton and baryon number conservation.
Moreover the simplifying assumption is made that a given leptoquark couples to 
just one family of fermions
which means that only first generation leptoquarks
can be produced in electron-photon scattering.
The first generation leptoquarks may decay into either an electron\footnote{Charge conjugation is implied throughout this paper for all particles, 
e.g.~positrons are also referred to as electrons.} and 
a quark or into a 
neutrino and a quark. The branching ratio of the decay into an electron
and a quark is commonly denoted by $\beta$.
Table~\ref{tab-lqstates} shows all the possible leptoquark states
considered in this paper
(scalar and vector) using the most common nomenclature \cite{bib-nom} along 
with their electric charge and fermion number $F=L+3B$.
The branching ratios $\beta$ given in this table assume that
the leptoquarks couple only to Standard Model particles.
To respect the existing limits on the product $\lambda_{\rm L}\cdot
\lambda_{\rm R}$ we assume that
for the scalar states S$_0$ and S$_{1/2}$ and the vector states
V$_0$ and V$_{1/2}$ either the left-handed or the right-handed
coupling must vanish, i.e. $\lambda_{\rm L}\cdot\lambda_{\rm R}=0$.

At an e$^+$e$^-$ collider different diagrams are expected to 
contribute to single leptoquark production in electron-photon 
collisions~\cite{bib-doncheski,bib-aliev,bib-london,bib-boos,bib-hewett}.
The dominant diagram is $\mbox{eq}\to\LQ$ (Fig.~\ref{fig-gamres}),
where a photon, which has been radiated by one of the beam electrons, serves
as a source of quarks through its fluctuations into hadronic states.
The electron-quark interaction produces a leptoquark which is 
assumed to decay subsequently into an electron or a neutrino, 
and a quark. The photon remnant may disappear down the beam-pipe 
or add some activity in the forward region of the detector.
The diagrams shown in Fig.~\ref{fig-gampoint} are also relevant, whereas
$\ee$ annihilation diagrams with a single leptoquark 
radiated in the final state
and diagrams with quark and/or leptoquark exchange in the t-channel but without
photon exchange are suppressed. 
The signature of single leptoquark events is
one hadronic jet with high transverse momentum, balanced
either by one isolated electron
or by missing transverse energy due to the neutrino.
Both topologies are studied in this paper.

Squarks ($\tilde{\mbox{q}}$)
in supersymmetric models with R-parity violation have the same 
production mechanism as some leptoquarks. 
R-parity is a quantum number which equals $+1$ for particles and $-1$ 
for their superpartners.
Table~\ref{tab-lqstates} shows the correspondence between the squark 
and the leptoquark states.
R-parity conserving decays are possible for squarks, in addition to the 
R-parity violating leptoquark decay modes.
The ratio between the R-parity
conserving and violating modes depends on the parameters of supersymmetry and
on the size of the coupling. 
For this analysis the branching ratio for R-parity
conserving decays has been set to zero and consequently $\beta$ 
has the same value for squarks and the corresponding leptoquarks.
Supersymmetry allows only left-handed couplings to leptons for these states. 
In the most general superpotential of the Minimal Supersymmetric Model (MSSM),
the renormalizable gauge invariant operator which describes the 
coupling of squarks to quarks and leptons and violates R-parity
is $\lambda'_{ijk}L^i_{\rm L}Q^j_{\rm L}\bar{D}^k_{\rm R}$~\cite{susyrpv},
where $i$, $j$ and $k$ are generation indices of the left-handed
doublet superfields of leptons ($L_{\rm L}$) and quarks ($Q_{\rm L}$), and 
right-handed singlets of down-type quarks ($\bar{D}_{\rm R}$).
Only couplings $\lambda_{1jk}'$ to first generation leptons are
considered in this paper. 

\begin{table}[htbp]
  \begin{center}
    \begin{tabular}{|c|c|c|c|c|}\hline   
  scalar LQ($\tilde{\mbox{q}}$) & charge & $F$ & decay mode  & $\beta$ \\ 
\hline
\rule[0.5cm]{0cm}{0cm}    \~S$_{\rm 0}$ (or ${\rm \tilde{d}}_R$) 
& -1/3 & 2 & e$^-_{\rm L}$u, $\nu_{\rm L}$d & 1/2 \\
  &   &      & e$^-_{\rm R}$u                 &  1   \\ \hline
\rule[0.5cm]{0cm}{0cm}    S$_{\rm 0}$ & -4/3 & 2 & e$^-_{\rm R}$d & 1 \\ \hline
\rule[0.5cm]{0cm}{0cm}  \~S$_{\rm 1/2}$ (or ${\rm \bar{\rm \tilde{d}}_L}$) 
& +1/3 & 0&  $\nu_{\rm L}\bar{\rm d}$ & 0 \\
\rule[0.5cm]{0cm}{0cm}  \~S$_{\rm 1/2}$ (or ${\rm \bar{\rm \tilde{u}}_L}$) & -2/3 & 0&  e$^-_{\rm L}\bar{\rm d}$ & 1 \\ \hline
\rule[0.5cm]{0cm}{0cm}  & +2/3  & &  $\nu_{\rm L}$u & 0 \\
    S$_1$               & -1/3  &2&  $\nu_{\rm L}$d, e$^-_{\rm L}$u & 1/2 \\
                        & -4/3  & &  e$^-_{\rm L}$d  & 1 \\ \hline
\rule[0.5cm]{0cm}{0cm}  & -2/3  & &  $\nu_{\rm L}\bar{\rm u}$ & 0 \\
 S$_{1/2}$              &       &0&e$^-_{\rm R}\bar{\rm d}$ & 1 \\
                        & -5/3  & &  e$^-_{\rm L}\bar{\rm u}$ or e$^-_{\rm R}\bar{\rm u}$ & 1 \\ \hline \hline
 vector LQ  & charge & $F$& decay mode  & $\beta$ \\ \hline
\rule[0.5cm]{0cm}{0cm} & -1/3 & & $\nu_{\rm L}$d & 0 \\
            V$_{1/2}$  &      & 2 &  e$^-_{\rm R}$u  & 1 \\
                       & -4/3 &   & e$^-_{\rm R}$d or e$^-_{\rm L}$d & 1 \\ \hline
\rule[0.5cm]{0cm}{0cm} \~V$_{1/2}$  & +2/3 &2& $\nu_{\rm L}$u & 0 \\
                                    & -1/3 & & e$^-_{\rm L}$u & 1 \\ \hline 
\rule[0.5cm]{0cm}{0cm} V$_0$ & -2/3&0 & e$^-_{\rm L}\bar{\rm d}$, $\nu_{\rm L}\bar{\rm u}$ & 1/2 \\
                             & &     & e$^-_{\rm R}\bar{\rm d}$ & 1 \\ \hline
\rule[0.5cm]{0cm}{0cm}  & +1/3 & & $\nu_{\rm L}\bar{\rm d}$ & 0 \\
   V$_1$                   & -2/3 &0 & e$^-_{\rm L}\bar{\rm d}$,  $\nu_{\rm L}\bar{\rm u}$ & 1/2 \\
               & -5/3 && e$^-_{\rm L}\bar{\rm u}$ & 1 \\ \hline
\rule[0.5cm]{0cm}{0cm}  \~V$_0$       & -5/3 &0& e$^-_{\rm R}\bar{\rm u}$ & 1 \\ \hline
  \end{tabular}
    \caption{All possible scalar (S) leptoquarks/squarks and vector (V) 
leptoquarks in the BRW model 
with their electric charge in units of $e$, their fermion
number $F$, their decay modes and the corresponding branching ratio $\beta$ 
for the decay into an electron and a quark.}
    \label{tab-lqstates}
  \end{center}
\end{table}

Several experiments have searched for leptoquarks. 
DELPHI analysed the single scalar and vector leptoquark production setting 
limits on the mass ranging from 134 GeV$/c^2$ to 171 GeV$/c^2$ at an 
$\ee$ centre-of-mass energy $\sqrt{s}_{\rm ee}$ of 
183 GeV~\cite{bib-delphi} and for a Yukawa coupling 
$\lambda>\sqrt{4\pi\alpha_{\rm em}}$.
The H1 collaboration at HERA has searched for leptoquarks in deep-inelastic
neutral current (NC) and charged current (CC)
electron-proton scattering at high $Q^2$~\cite{bib-h1lq}. 
First generation scalar (vector) leptoquarks have been excluded for masses
up to 275~GeV$/c^2$ (284~GeV$/c^2$) for $\lambda>\sqrt{4\pi\alpha_{\rm em}}$
and fermion number $F=0$, while masses up to about 200~GeV$/c^2$
have been excluded for $|F|=2$ leptoquarks. Leptoquark
limits as a function of the assumed couplings 
have also been obtained from fermion pair production
at LEP2 at $\sqrt{s}_{\rm ee}=130-183$~GeV~\cite{bib-fpair}.

Leptoquark pair production limits have been 
obtained by the LEP experiments at 
$\sqrt{s}_{\rm ee}=M_{\rm Z}$~\cite{bib-pair1} and at 
$\sqrt{s}_{\rm ee}=183$~GeV~\cite{bib-pair2}, 
and by the D0~\cite{bib-d0} and CDF experiments~\cite{bib-cdf}
for leptoquarks of the three generations.
The leptoquark pair production limits are independent of
the Yukawa coupling $\lambda$ in $\ppbar$ scattering. 
In $\ee$ scattering the leptoquark pair production cross-section
can be considered independent of $\lambda$ only in the region
of small $\lambda$ where t-channel quark exchange can be neglected 
compared to the s-channel diagram. 
At LEP2 energies, $\sqrt{s}_{\rm ee}=183$~GeV, the
mass limits for first generation leptoquarks vary between 80~GeV$/c^2$
and 90~GeV$/c^2$, depending on the leptoquark state~\cite{bib-pair2}.
Scalar leptoquarks of charge $-1/3$ are only excluded for 
$M_{\rm LQ}<M_{\rm Z}/2$~\cite{bib-pair1}.
A combination of the CDF and D0 search results for a first
generation scalar leptoquark yields a lower mass limit
of 242~GeV$/c^2$ for $\beta=1$~\cite{bib-cdfd0}.
The CDF and D0 collaborations~\cite{bib-squarks} 
as well as the four LEP experiments~\cite{lepsquarks} 
have also searched for pair production of R-parity 
violating squarks.

Leptoquark pair production limits obtained at LEP
are sensitive to the mass region $M_{\rm LQ}<\sqrt{s_{\rm ee}}/2$, whereas 
single leptoquarks can be produced almost up to the kinematic limit,
$M_{\rm LQ}=\sqrt{s_{\rm ee}}$. Even though the leptoquark mass range covered
by CDF and D0 in $\bar{\mbox{p}}$p scattering is higher for most 
leptoquark states, the analysis presented here 
is more sensitive in the low $\beta$ region ($\beta\to 0)$. 
For $\beta\equiv 0$ no production in eq collisions is possible. 
In addition,
this search is also sensitive to the production of ec (and es) states,
but only flavour-diagonal couplings $\lambda$ are considered in this paper.

We present a search for leptoquarks with $M_{\rm LQ}>80$~GeV
in electron-photon scattering using
data corresponding to an integrated luminosity of 
$164.7$~pb$^{-1}$ (eq channel) and $158.4$~pb$^{-1}$ ($\nu$q channel)
at e$^+$e$^-$ centre-of-mass energies of 189 GeV. 

\section{The OPAL detector}
\label{sec-opal}  
The OPAL detector is described in detail in~\cite{opaltechnicalpaper}. 
It is a multipurpose apparatus
having nearly complete solid angle coverage with excellent hermeticity.
The central detector consists of two layers of
silicon micro-strip detectors \cite{simvtx} surrounding the beam-pipe and
a system of gaseous tracking chambers inside a 0.435~T solenoidal magnetic field.

The lead-glass electromagnetic calorimeter (ECAL)
with a presampler is located outside the magnet coil. It provides,
in combination with the forward calorimeters (FD),
the forward scintillating tile counter (the ``MIP plug'')~\cite{bib-llpaper},
and the silicon-tungsten luminometer (SW)~\cite{bib-siw}, 
a geometrical acceptance
down to 25~mrad from the beam direction.  The SW luminometer
measures the integrated luminosity using small-angle Bhabha
scattering events~\cite{lumino}.
The magnet return yoke is instrumented for hadron calorimetry (HCAL).
It is surrounded by several layers of muon chambers.

\section{Kinematics and Monte Carlo simulations}
\label{sec-mc}
The Monte Carlo simulation of the process $\mbox{e}^+\mbox{e}^-\to \LQ +\X$
is done with the program ERATO-LQ~\cite{bib-erato} which can generate all 
the states listed 
in Table~\ref{tab-lqstates} and calculates the cross-sections for
the scalar and vector states\footnote{All total cross-sections
in this paper are defined as a sum of the particle and the anti-particle
state.}.

\begin{figure}[ht]
   \begin{center}
          \epsfxsize=10cm
          \epsffile{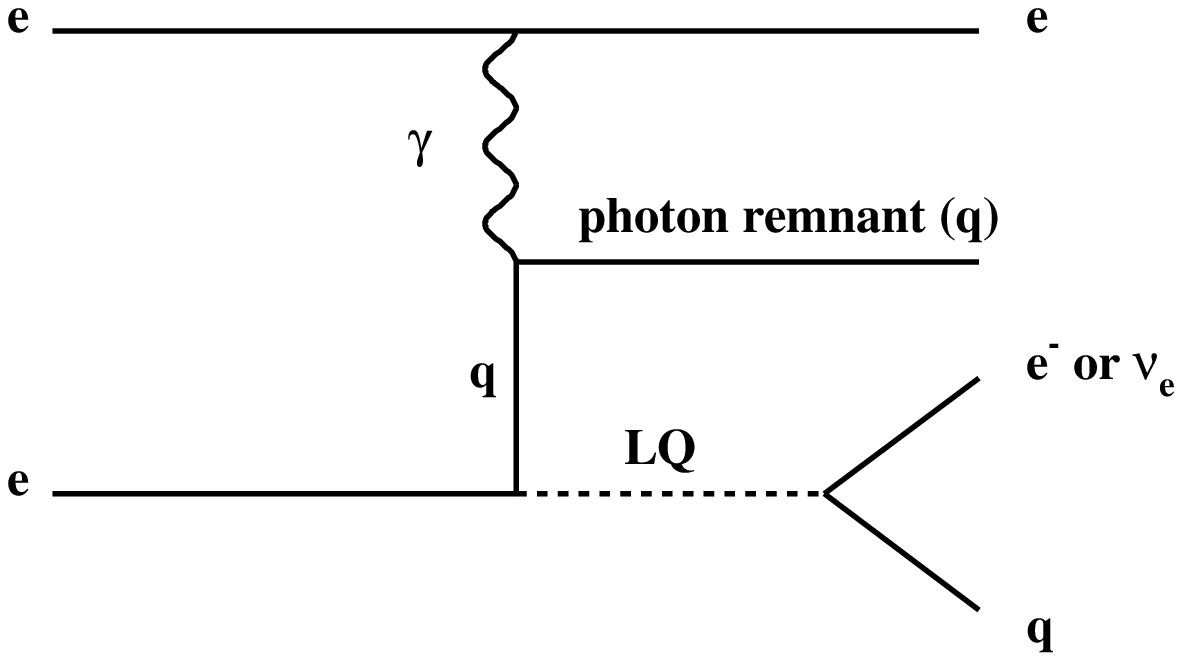}
\caption{Diagram of the $s$-channel
production of a leptoquark (LQ) in electron-photon
scattering. The photon is radiated by one of the LEP beams, fluctuates into 
a hadronic object and one of the quarks interacts with an electron from 
the other beam.}
\label{fig-gamres}
\vspace{0.5cm}
%
          \epsfxsize=10cm
          \epsffile{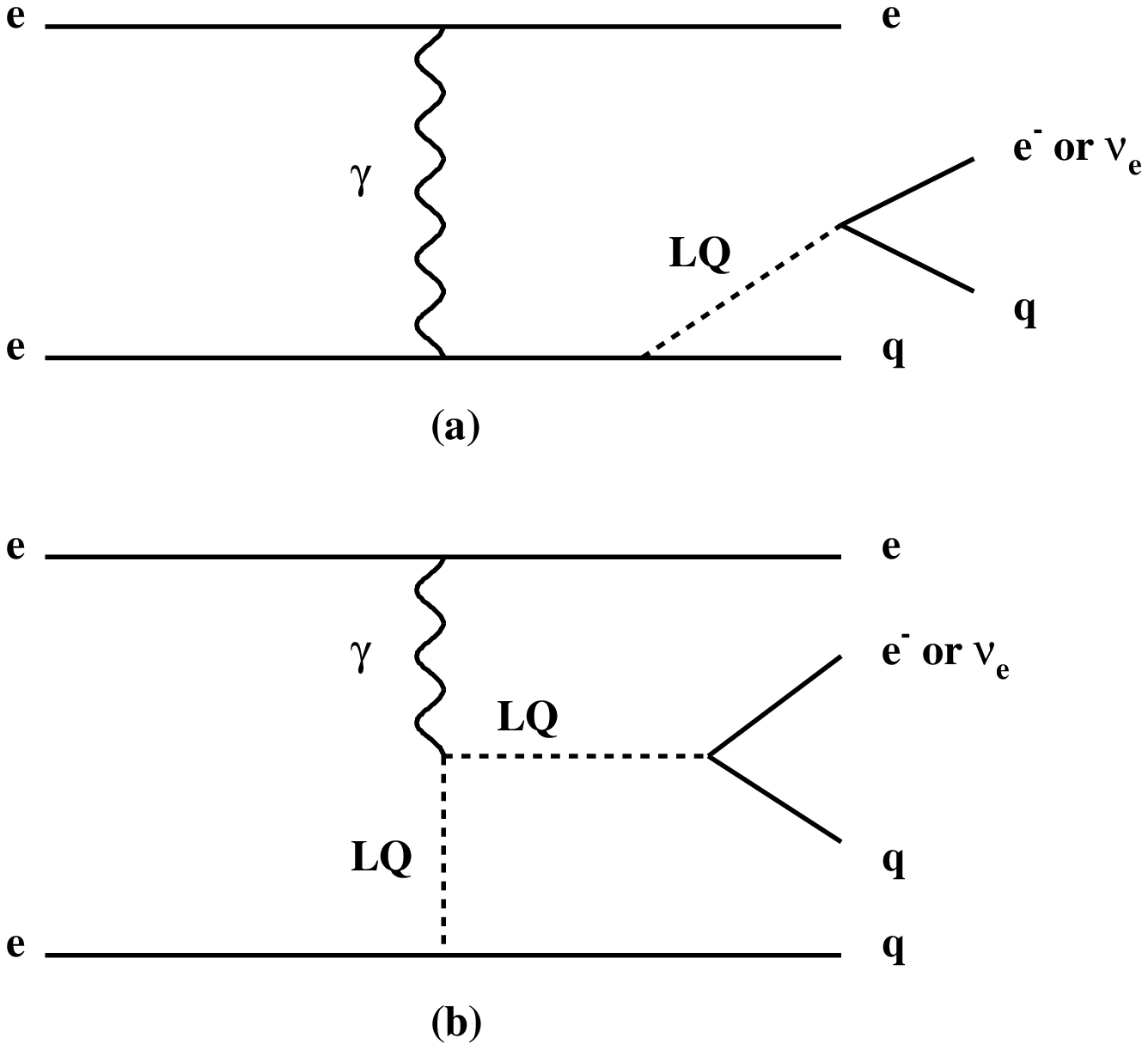}
\caption{Alternative single leptoquark production mechanisms, where the 
photon interacts pointlike: 
({\bf a}) ``direct'' interaction with the beam electron, 
({\bf b}) photon is absorbed by the leptoquark ``emitted'' by the other beam electron.}
\label{fig-gampoint}
\end{center}
\end{figure}

The total cross-section for the production of leptoquarks
of mass $M_{\rm LQ}$ can be written as 
a convolution of the probability to find a photon with the
momentum fraction $z$ in the electron,
approximated here by the Weizs\"acker-Williams effective photon
distribution $\fg(z)$~\cite{bib-wwa},
and the probability to find a quark in the photon. This probability can 
be parametrised
by parton distribution functions (pdf) $f_{{\rm q}/\gamma}(x,\mu^2)$ 
of the photon, which are evaluated at
the scale $\mu^2=M^2$~\cite{bib-doncheski}.
The Bjorken scaling variable $x$ is given by
$x=M^2/zs$. With these assumptions the total cross-section
for scalar leptoquark production is:
\begin{equation}
\sigma(\ee\rightarrow\LQ + \X) =\frac{\lambda^2\pi}{2s}
\int_{M^2/s}^1\frac{dz}{z}f_{\gamma/e}(z)
f_{{\rm q}/\gamma}(M^2/(zs),M^2)
\end{equation}
where $f_{{\rm q}/\gamma}(M^2/(zs),M^2)$ is obtained by convoluting 
the parton level 
cross section with the quark distribution $f_{{\rm q}/\gamma}(z,M^2)$ 
in the photon.
In case of unpolarised electron
beams, the total cross-section for the production of vector leptoquarks is twice
as large as the cross-section for scalar leptoquarks~\cite{bib-buch}.
This approach based on the pdf is used in the calculations 
of the diagram shown in Fig.~\ref{fig-gamres} by 
Donchesky~\cite{bib-doncheski} and it is also implemented in the Monte Carlo 
generator PYTHIA~\cite{bib-pythia}.

However, PYTHIA can only generate scalar leptoquarks with charge 
$-1/3$. It has therefore only been used to check the 
cross-sections calculated by ERATO-LQ~\cite{bib-erato}. 
This Monte Carlo generator uses a perturbative calculation of
the diagrams in Fig.~\ref{fig-gamres} and ~\ref{fig-gampoint}.
It is expected to give the correct angular distributions
both for the scalar and vector leptoquarks.

Cross-sections calculated with ERATO-LQ and 
PYTHIA using $\lambda=\sqrt{4\pi\alpha_{\rm em}}$ and charge $-1/3$
(e.g for S$_{0}$) are shown in Table~\ref{tab-cross}. 
Using a different parametrisation of the pdf, GRV~\cite{bib-grv}
instead of SaS-1D~\cite{bib-sas},
has almost no effect on the resulting cross-section. 
In ERATO-LQ the exact total cross-section given in~\cite{bib-london},
taking into account all possible diagrams, can also be calculated.
In comparison, the total cross-section calculated perturbatively by the MC
generator ERATO-LQ is $10-20\%$ smaller than the exact total cross-section  
used to calculate the limits.

\begin{table}[htbp]
  \begin{center}
    \begin{tabular}{|c|c|c|c|c|c|c|c|c|} \hline   
&  $M_{\rm LQ}$~[GeV$/c^2$]   
& $80$     & $100$ & $120$ & $140$ & $160$ & $170$ & $180$ \\
 \hline
PYTHIA & $\sigma_{\rm tot}$~[pb]& $2.78$ & $1.31$ & $0.64$ & $0.32$ & 
$0.14$ & $0.083$ & $0.046$ \\ (GRV)    & & & & & & & & \\ \hline 
PYTHIA & $\sigma_{\rm tot}$~[pb]& $2.77$ & $1.30$ & $0.64$ & $0.32$ & 
$0.14$ & $0.086$ & $0.046$ \\ (SaS-1D)    & & & & & & & & \\ \hline
ERATO-LQ~\cite{bib-london} & $\sigma_{\rm tot}$~[pb] & $2.81$ & $1.46$ & 
$0.77$ & $0.40$ & 
$0.18$ & $0.102$ & $0.037$ \\ (exact)    & & & & & & & & \\ \hline
ERATO-LQ & $\sigma_{\rm tot}$~[pb]    & $2.32$ & $1.21$ & 
$0.65$ & $0.34$ & 
$0.16$ & $0.085$ & $0.031$ \\ (perturb.) & & & & & & & & \\  \hline
    \end{tabular}
    \caption{
The total cross-section for the single production of the state S$_{0}$
with charge $-1/3$ as a function of the leptoquark mass $M_{\rm LQ}$
calculated by the Monte Carlo generators ERATO-LQ and PYTHIA
at $\sqrt{s}_{\rm ee}=189$~GeV using $\lambda=\sqrt{4\pi\alpha_{\rm em}}$.
}
    \label{tab-cross}
  \end{center}
\end{table}

ERATO-LQ generates 
the four-vectors of the direct decay products of the leptoquark.
A scalar leptoquark decays isotropically in its rest frame leading to
a flat distribution in the variable $y=(1+\cos\theta^*)$, where
$\theta^*$ is the decay angle of the lepton relative to
the incident quark in the leptoquark centre-of-mass frame. 
The decay angles in the decays of vector leptoquarks are distributed according
to ${\rm d}\sigma/{\rm d}y\propto (1-y)^2$.
The simulated photon is always real, i.e., the negative squared
four-momentum of the photon, $Q^2$, is identical to zero.

The partial decay widths of a scalar (S) and a vector (V) leptoquark are 
\begin{equation}
\Gamma_{\rm S}=\frac{3}{2}\Gamma_{\rm V}=\frac{\lambda^2}{16\pi} M_{\rm LQ}.
\end{equation}
Since the leptoquark carries colour, it could hadronise
before its decay into fermions. This effect is taken into
account in the systematic uncertainties but not in the 
standard Monte Carlo simulation, since
it should only be important for $\Gamma_{\rm S,V}\ll\Lambda_{\rm QCD}$. 
The decay width $\Gamma_{\rm S}$ is
$16$~MeV for $\lambda=0.1$ and $M_{\rm LQ}=80$~GeV$/c^2$.

JETSET~\cite{bib-pythia} is used to perform the hadronisation
of the leptoquark decay products.
It has been checked with ERATO-LQ that the 
event properties of the different leptoquark states are very similar 
which allows to simplify the generation considerably: 
For each of the seven masses 
listed in Table~\ref{tab-cross}, samples of 3000 events for scalar and 
vector leptoquarks were generated for the two leptoquark decay modes separately
and only for one state.
Also, no extra squark events needed to be generated.

All relevant Standard Model background processes have been studied using Monte Carlo
generators. Multi-hadronic events ($\ee\rightarrow \qqbar(\gamma)$) have been
simulated with PYTHIA 5.722 \cite{bib-pythia}.
KORALZ 4.02 \cite{bib-koralz} has been used
to generate the process $\ee\rightarrow\tau^+\tau^-$ and
BHWIDE \cite{bib-bhwide} to generate the Bhabha process $\ee\rightarrow\ee$.

Deep inelastic e$\gamma$ events in the
range $Q^2>4.5$~GeV$^2$ including charged current 
deep inelastic scattering (CC DIS) events have been simulated with 
HERWIG 5.8 \cite{bib-herwig}.
PHOJET 1.10~\cite{bib-phojet} has been used to generate hadronic two-photon events
(i.e. $\ee\rightarrow\ee\mbox{~hadrons}$) in the range $Q^2<4.5$~GeV$^2$.
Leptonic two-photon events have been generated with
Vermaseren~\cite{bib-vermaseren}. The events generated with HERWIG, 
PHOJET and Vermaseren are called two-photon events. An alternative
hadronic two-photon sample for systematic studies has been generated using 
F2GEN~\cite{bib-f2gen} for $Q^2>4.5$~GeV$^2$ and PYTHIA for
$Q^2<4.5$~GeV$^2$.
Other processes with four fermions in the final states, including 
W pair production, have been simulated with grc4f~\cite{bib-grc4f}
and an alternative sample has been generated with KORALW~\cite{bib-koralw}.
The generated signal events and all the background
Monte Carlo events have been passed through a full detector 
simulation and the same reconstruction algorithms as the real data.

\section{Event analysis}
\label{sec-evsel}
We search for events with one hadronic jet and either an 
electron or missing energy balancing the transverse momentum of this jet.
The analysis uses tracks measured in the central tracking devices,
clusters measured in the ECAL, the HCAL, the FD and the SW.
In addition to quality requirements which ensure that 
the tracks have their origin close to the $\ee$ interaction
point, tracks must have more than 20 hits in the central jet chamber
and more than half the number of hits expected for the given track.
The transverse momentum of the track with respect to
the beam direction must be greater than 120 MeV.
Tracks with a momentum error larger than the momentum itself
are rejected if they have fewer than 80 hits. 
Calorimeter clusters have to pass energy threshold cuts to
suppress noise.
To avoid double counting of particle momenta, a matching algorithm
between tracks and clusters is applied~\cite{bib-cone}.
Clusters are rejected if the energy of the cluster is 
less than expected from the momentum of an associated track.
If the cluster energy exceeds
the expected energy by more than what is expected from the resolution,
the expected energy is subtracted from the cluster energy.
In this case the track momentum 
and the reduced energy of the cluster are counted separately.

The tracks and remaining clusters are used as input to the jet 
finding algorithm and to determine
the missing transverse energy of the event.
Jets are reconstructed using
a cone jet finding algorithm with a cone size $R=1$ and 
a minimum transverse jet energy $E_{\rm T}$ of
15~GeV~\cite{bib-cone}. 
The cone size $R$ is defined as
$R=\sqrt{(\Delta\eta)^2+(\Delta\phi)^2}$,
with $\eta=-\ln\tan(\theta/2)$ being the pseudorapidity, $\phi$ the
azimuthal angle and $\theta$ the polar angle in the laboratory frame 
in radians. 
$\Delta\eta$ and $\Delta\phi$ are the differences
between the cone axis and the particle direction.
No cut on the pseudorapidity of the jet is used at this stage.

\subsection{The electron plus hadronic jet channel}
\begin{figure}[htbp]
   \begin{center}
      \makebox[15cm][r]{
          \epsfxsize=15cm
          \epsffile{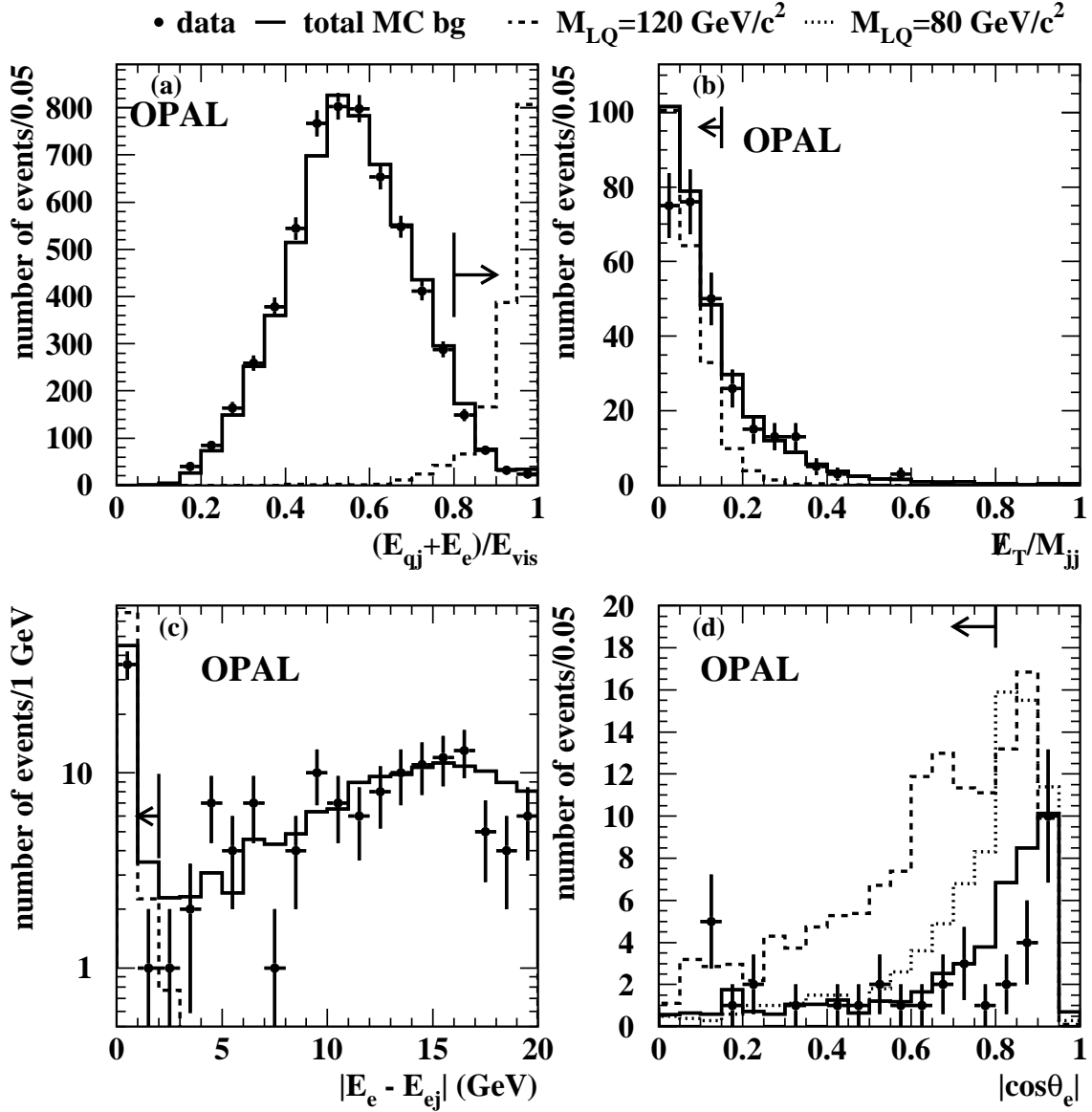}
           }
   \end{center}
\caption{Electron-quark decay channel:
({\bf a}) distribution of the ratio 
$(E_{\rm qj}+E_{\rm e})/E_{\rm vis}$ after the 
preselection;
({\bf b}) distribution of the ratio $\ETMISS/M_{\rm jj}$ 
after applying the additional cut EQ1;
({\bf c}) distribution of the difference $|E_{\rm e}-E_{\rm ej}|$
after applying the additional cut EQ2;
({\bf d}) distribution of the electron scattering angle 
$|\cos\theta_{\rm e}|$ after  
applying cuts EQ1 through EQ3 to the preselected data.
The points with error bars are the data, the full line represents the total 
Standard Model background normalised to data luminosity and the dashed (dotted)
histogram shows the distribution for the vector leptoquark 
state \~V$_{0}$ with a mass of 120 GeV$/c^2$ (80 GeV$/c^2$).
The normalisation of the leptoquark signals is arbitrary.}
\label{fig-cutseq1}
\end{figure}

In this search the selection cuts were optimised for a leptoquark
decaying into a single quark and an isolated electron. 
The electron is identified by requiring a minimum of
20 hits used in the measurement of the specific energy loss,
${\mathrm d}E/{\mathrm d}x$, and a ${\mathrm d}E/{\mathrm d}x$ probability 
for the electron hypothesis of more than $1\%$.
Furthermore, the ratio of the total energy of the electron measured
in the ECAL to the momentum of the track
associated to this ECAL cluster must lie between $0.7$ and $2$. 
The identified electron with
the largest momentum was assumed to be the electron from the leptoquark
decay. 
Since the electron is included in the jet search, it is
usually reconstructed as a jet.
Candidate leptoquark events are selected based on
the following cuts, which are identical for scalar and vector leptoquarks:
\begin{itemize}
\item
The event must contain more than four tracks ($n_{\rm ch}>4$).
\item
Exactly two jets must have been reconstructed ($n_{\rm jet}=2$). 
One of the jets must contain the highest energy electron. 
\item 
The highest energy electron must have an energy $E_{\rm e}$ 
greater than 2~GeV. The electron energy $E_{\rm e}$ is the energy
of the calorimeter clusters matched to the electron track.
This cut is effective against all sources of Standard Model background, 
especially two-photon events.  
\item
The jet not containing the electron must consist of more than six particles
($n_{\rm qj}>6$),
where the number $n_{\rm qj}$ of particles is defined as the sum of the
number of tracks and calorimeter clusters after matching. 
This cut reduces the number of $\ee\to\tau^+\tau^-$ and
the number of Bhabha events and also some of the remaining leptonic 
two-photon events.
\item The total energy $E_{\rm HCAL}$ measured in the hadronic calorimeter 
has to be greater than 1~GeV. This cut is effective against Bhabha events. 
\end{itemize}
After this preselection, 5739 data events remain. The selection
efficiencies are given in Table~\ref{tab-events1a}. They
are significantly smaller for the vector leptoquark states than
for the scalar leptoquark states
due to the angular distribution of the decay electrons which
is peaked at $\cos\theta=\pm 1$ for vector leptoquark states.
The following set
of cuts is applied to the remaining events: 
\begin{itemize}
\item[EQ1)] 
To ensure that most of the measured energy comes from the two jets,  
the energy $E_{\rm qj}$ of the hadronic jet and the electron energy 
$E_{\rm e}$ must add up to more than $80\%$ of the visible energy 
$E_{\rm vis}$, i.e.~$\left((E_{\rm qj}+ E_{\rm e})/E_{\rm vis}>0.8\right)$. 
This cut is efficient against multihadronic and four-fermion 
events (Fig.~\ref{fig-cutseq1}a).
\item[EQ2)] 
The ratio of the missing transverse energy $\ETMISS$ and the invariant 
mass $M_{\rm jj}$ of the two jets has to be 
$\ETMISS/M_{\rm jj}<0.15$. The invariant mass $M_{\rm jj}$ of the two
jets is calculated from the four-vectors of the two reconstructed jets. 
This cut reduces the four-fermion background by a factor 
two (Fig.~\ref{fig-cutseq1}b).
\item[EQ3)] 
Since the electron is expected to be isolated, 
the difference between the electron energy E$_{\rm e}$ and the 
energy E$_{\rm ej}$ of the jet which contains the electron has to 
be smaller than 2~GeV ($|\rm E_{\rm e}-E_{\rm ej}|<2$ GeV). 
This eliminates most of the remaining $\ee\to\qqbar(\gamma)$ 
events (Fig.~\ref{fig-cutseq1}c).
\item[EQ4)] 
The electron must lie in the angular region defined by 
$|\cos\theta_{\rm e}|<0.8$ (Fig.~\ref{fig-cutseq1}d).  
This cut rejects mainly 
two-photon events with a scattered electron within the detector
acceptance. 
\end{itemize}
In Table~\ref{tab-events1a} the number
of data events and the expected number of Standard Model background events 
taken
from the Monte Carlo are shown after the preselection and after each
subsequent cut.
The number of Monte Carlo events has been normalised to the data luminosity.
The selection efficiencies for three different scalar 
and vector leptoquark masses are also given. 
\renewcommand{\arraystretch}{1.0}
\begin{table}[htbp]
  \begin{center}
    \begin{tabular}{|c|c|c|c|c|c|c|c|}    
\hline
\multicolumn{2}{|c|}{cuts} 
& Pre-$\pzz$ & 
$\frac{\rm  E_{\rm qj}+ E_{\rm e}}{\rm  E_{\rm vis}}$ & $\ETMISS/M_{\rm jj}$ 
& $|\rm E_{\rm e}-E_{\rm ej}|$ & $|\cos\theta_{\rm e}|$ \\ \cline{1-2}
\rule[0.5cm]{0cm}{0cm}  
$M_{\rm LQ}$~[GeV$/c^2$] & state & selection & (EQ1) & (EQ2) & (EQ3) & (EQ4) \\
 \hline \hline
$\pz80$ & & 49.6\% & 44.7\% & 39.4\% & 37.9\% & 28.4\% \\ 
\cline{1-1}\cline{3-7}
$120$ & scalar & 61.4\% & 58.7\% & 54.1\% & 51.6\% & 45.1\% \\ 
\cline{1-1}\cline{3-7}
$160$ &  & 70.7\% & 70.5\% & 65.4\% & 60.2\% & 55.0\% \\ \hline \hline
$\pz80$& & 37.0\% & 31.2\% & 27.8\% & 26.7\% & 12.3\% \\
\cline{1-1}\cline{3-7}
$120$ & vector & 50.4\% & 47.5\% & 43.9\% & 41.6\% & 29.4\% \\ 
\cline{1-1} \cline{3-7}
$160$ &  & 61.5\% & 61.3\% & 57.4\% & 54.0\% & 44.2\% \\ 
\hline \hline 
\multicolumn{2}{|c|}{$\ee\rightarrow \qqbar$} & 4834 & 200.1 & 160.1 & $\pz 2.7$ & 
$\pz2.2 \pm 0.2$ \\ \hline
\multicolumn{2}{|c|}{$\ee\rightarrow\tau^+\tau^-$} & 13.1 & 5.2 & 2.7 & $\pz 0.7$ & 
$\pz0.5\pm0.1$ \\ \hline
\multicolumn{2}{|c|}{$\ee\rightarrow\ee$} & 0.5 & 0.5 & 0.5 & $\pz 0.5$ & $\pz0.2 \pm 0.1$ \\ \hline
\multicolumn{2}{|c|}{$\ee\rightarrow 4\mbox{~fermions}$} & 817.2 & 72.8 & 33.0 & 18.2 & $12.8\pm 0.5$ \\ \hline
\multicolumn{2}{|c|}{two-photon} & 62.2 & 30.0 & 26.2 & 25.5 & 
$\pz6.3\pm 1.0$ \\ \hline\hline
\multicolumn{2}{|c|}{total BG} & 5727 & 308.5 & 222.4 & 47.5 & $21.9\pm 1.1$ \\ \hline
\multicolumn{2}{|c|}{data}  & 5739 & 270$\pz$ & 194$\pz$ & 36$\;\pz$ & 21$\pz\pzz$ \\ \hline
    \end{tabular}
    \caption{Selection efficiencies for three different leptoquark masses for 
scalar and vector leptoquarks. The remaining number of data
 events and the expected number of background (BG) events are also
listed after each cut of the electron
plus hadronic jet selection. The Monte Carlo background is normalised to
the data luminosity of $164.7$~pb$^{-1}$ at $189$~GeV. 
The errors on the Monte Carlo background are statistical.}
    \label{tab-events1a}
  \end{center}
\end{table}

Figs.~\ref{fig-cutseq1}a-d show the distributions of some of 
the cut variables for data, Standard Model background and the leptoquark state 
\~V$_{0}$ with a mass of 120 GeV. The data distributions
are in general well described by the Monte Carlo simulation.
After all cuts the Standard Model background is expected
to be mainly due to four-fermion and two-photon interactions.

The $|\cos\theta_{\rm e}|$ distribution 
depends strongly on the leptoquark mass. This is 
shown in Fig.~\ref{fig-cutseq1}d where an additional distribution 
for a leptoquark mass of 80~GeV has been added.
The cut $|\cos\theta_{\rm e}|<0.8$ is necessary to reduce the
background from two-photon events, but it also significantly reduces
the efficiency for small $M_{\rm LQ}$.

\begin{figure}[htbp]
   \begin{center}
          \epsfxsize=14cm
          \epsffile{pr337_04.epsi}
   \end{center}
\caption{({\bf a}) Selection efficiency in the eq channel 
after all cuts for
scalar (dots) and vector (squares) leptoquarks;
({\bf b}) 
Invariant mass $M_{\rm jj}$
of the two jets (= leptoquark mass) after all cuts for data (points
with error bars), 
Standard Model background (full line) and two different vector leptoquark masses 
(dotted line for 80 GeV$/c^2$ and dashed line for 120 GeV$/c^2$). 
The state \~V$_{0}$ was chosen to normalise the signal, 
using $\lambda=\sqrt{4\pi\alpha_{\rm em}}$.}
\label{fig-meqvector}
%
   \begin{center}
          \epsfxsize=14cm
          \epsffile{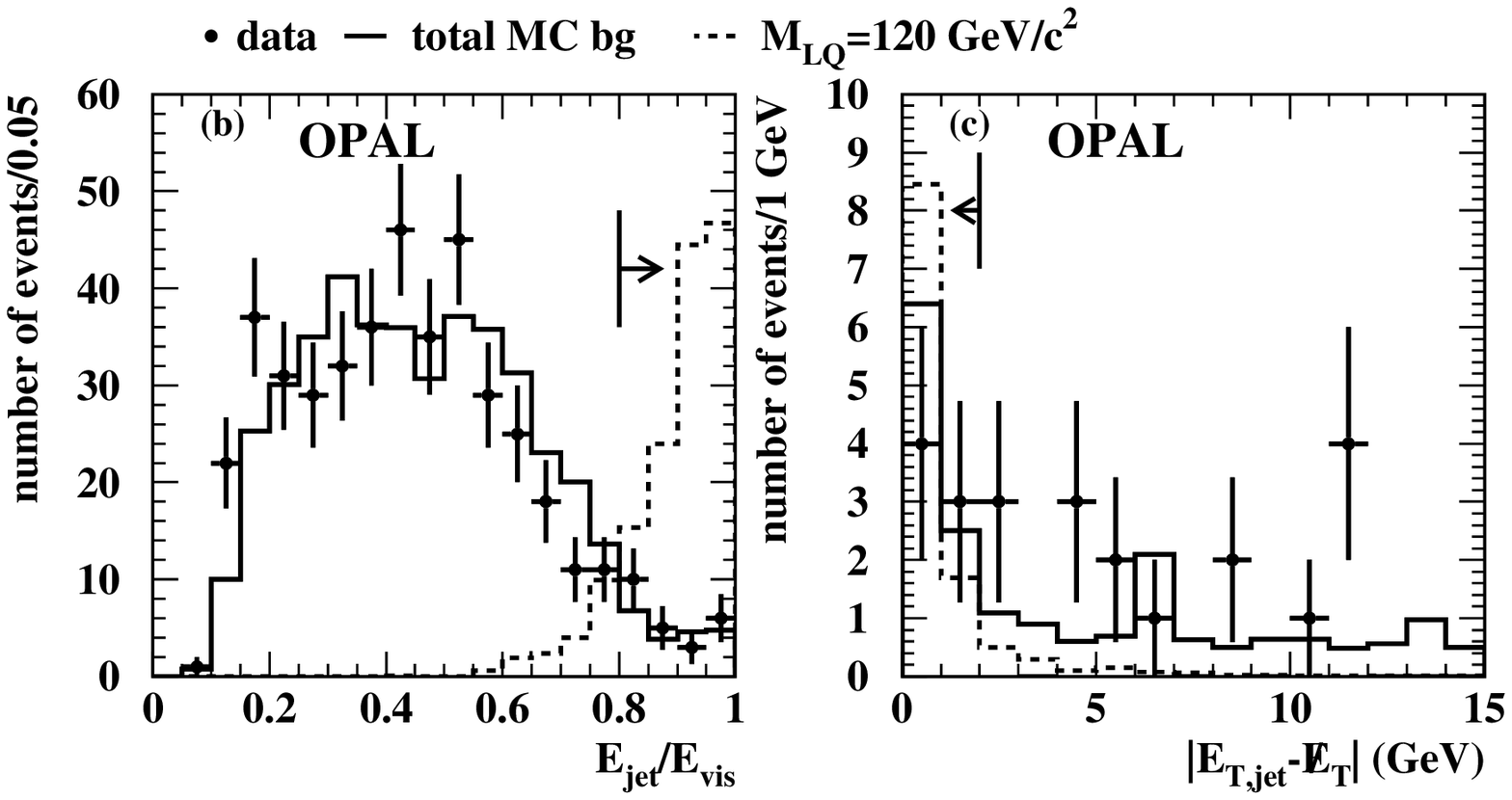}
   \end{center}
\caption{Neutrino-quark decay channel:
({\bf a}) distribution of the ratio $E_{\rm jet}/E_{\rm vis}$ 
after the preselection;
({\bf b}) distribution of the variable $|E_{\rm jet}-\ETMISS|$ after the 
cut NQ1 $(\left(E_{\rm jet}/E_{\rm vis}>0.8\right)$.
The points with error bars are the data, the full line represents the total 
Standard Model background normalised to data luminosity and the dashed 
histogram shows the distribution for the scalar state S$_{1/2}$ with a mass 
of 120 GeV$/c^2$.
The normalisation of the leptoquark signals is arbitrary.}
\label{fig-cutsnq1}
\end{figure}

Fig.~\ref{fig-meqvector}a shows the selection efficiencies after all 
cuts as determined with ERATO-LQ as a function of the generated 
leptoquark mass $M_{\rm LQ}$.
Fig.~\ref{fig-meqvector}b shows the distributions of $M_{\rm jj}$ for data, 
Standard Model background and for the state \~V$_{0}$ using 
$\lambda=\sqrt{4\pi\alpha_{\rm em}}$ with $M_{\rm LQ}=80$~GeV$/c^2$ and 
$120$~GeV$/c^2$.
After all cuts, 21 events remain in the data which is in good agreement 
with the predicted $21.9\pm1.1$~(stat) Standard Model background events.

\subsection{The neutrino plus hadronic jet channel}
In the case of the decay of a leptoquark 
into a neutrino and a single quark, the search has to be
optimised for a single hadronic jet in the detector. Its
transverse energy  $E_{\rm T,jet}$ must be balanced by the neutrino.
The cuts are therefore:
\begin{itemize}
\item
The event must contain more than four tracks 
($n_{\rm ch}>4$). 
\item
Exactly one jet must have been reconstructed in the pseudorapidity region 
$|\eta_{\rm j}|<2$. 
\item
No hit in the MIP plug with a significant charge deposition is found. 
\item
The jet must consist of more than six particles
($n_{\rm qj}>6$).
\item
The distance of the primary vertex to the nominal interaction point  
has to be less than 2~cm.
\end{itemize}
The following cuts are applied to the 432 data events which
remain after this preselection:
\begin{itemize}
\item[NQ1)] 
The ratio between the jet energy $E_{\rm jet}$ and the total visible 
energy $E_{\rm vis}$ in the event has to be greater than 0.8 
$\left(E_{\rm jet}/E_{\rm vis}>0.8\right)$. This cut 
is very effective in reducing
all sources of Standard Model 
background like multihadronic and two-photon events.
\item[NQ2)] 
The difference between the jet transverse energy, $E_{\rm T,jet}$, 
and the missing transverse energy,
$\ETMISS$, has to be less than 2~GeV ($|E_{\rm T,jet}-\ETMISS|<2$~GeV)
in order to ensure that the final state consists of a single jet
balanced by missing transverse energy.
\end{itemize}
\begin{table}[htb]
  \begin{center}
    \begin{tabular}{|c|c|c|c|c|}    
  \hline
\multicolumn{2}{|c|}{cuts} & Pre-$\pzz$ & 
$E_{\rm jet}/E_{\rm vis}$ & $|\rm E_{\rm T,jet}-\ETMISS|$ \\ 
\cline{1-2}
\rule[0.5cm]{0cm}{0cm} 
$M_{\rm LQ}$ (GeV$/c^2$) & state & selection & (NQ1) & (NQ2) \\ \hline \hline
$\pz80$ &  & 57.0\% &  45.4\% & 41.5\% \\ 
\cline{1-1} \cline{3-5}
$120$ & scalar & 62.3\% &  54.4\% & 48.3\% \\ 
\cline{1-1} \cline{3-5}
$160$ & & 67.7\% &  65.6\% & 56.1\% \\ \hline \hline
$\pz80$ &   & 59.0\% &  43.6\% & 39.9\% \\ 
\cline{1-1} \cline{3-5}
$120$ & vector & 59.2\% &  50.0\% & 43.7\% \\
\cline{1-1} \cline{3-5}
$160$ &  & 62.2\% &  60.1\% & 51.6\% \\ \hline \hline
\multicolumn{2}{|c|}{$\ee\rightarrow \qqbar$}
     & $178.2$  & $\pz2.9$ & $0.0\pm0.6$ \\ \hline
\multicolumn{2}{|c|}{$\ee\rightarrow\tau^+\tau^-$}
 &$\pzz4.6$ & $\pz2.1$ & $0.3\pm0.1$ \\ \hline
\multicolumn{2}{|c|}{$\ee\rightarrow 4\mbox{~fermions}$} 
                              &$\pz71.0$ & $10.5$   & $6.3\pm0.3$ \\ \hline
\multicolumn{2}{|c|}{two-photon} & $172.4$ &  $\pz4.6$ & $2.4\pm1.1$ \\ \hline \hline
\multicolumn{2}{|c|}{total BG}  & $426.2$ & $20.1$    & $8.9\pm1.2$ \\ \hline
\multicolumn{2}{|c|}{data}      & $432$ &  $24$    & $7$  \\ \hline
    \end{tabular}
    \caption{Selection efficiencies for three different leptoquark masses for 
scalar and vector leptoquarks. Also shown is the remaining number of data events and the expected
number of background (BG) events after each cut of the neutrino
plus hadronic jet selection. The Monte Carlo background is normalised to
the data luminosity of $158.4$~pb$^{-1}$ at $189$~GeV.
The errors on the Monte Carlo background are statistical.}
    \label{tab-selnuq}
  \end{center}
\end{table}
Table~\ref{tab-selnuq} shows the number of data and Monte Carlo 
events normalised to data luminosity after each cut beginning after the 
preselection. The signal efficiencies for 
three different scalar and vector leptoquark masses are also given.
The efficiencies are similar for scalar and for vector
states.

After all cuts, $7$ events remain which is in good agreement with the  
Standard Model expectation of $8.9\pm1.2$~(stat) events. 
From the two-photon events only 
the CC DIS events give a sizeable contribution 
to the final background composition together with the hadronic four-fermion 
processes. 

Fig.~\ref{fig-cutsnq1} shows some of the cut variables for data, 
Standard Model 
background and the leptoquark state \~S$_{\rm 1/2}$ with a mass of 
120~GeV$/c^2$ in arbitrary normalisation. 
The sum of the Standard Model Monte Carlo distributions 
describes the data sufficiently well.
Fig.~\ref{fig-mnuqscalar}a) shows the selection efficiencies as 
determined with the scalar and vector leptoquarks generated with ERATO-LQ 
after all cuts as a function of the generated leptoquark mass $M_{\rm LQ}$.
Fig.~\ref{fig-mnuqscalar}b) shows the distribution of the transverse
mass $M_{\rm T}=2\ETMISS$ after all cuts for data, 
Standard Model background, and for the state S$_{\rm 1/2}$ 
using $\lambda=\sqrt{4\pi\alpha_{\rm em}}$ with $M_{\rm LQ}=80$~GeV$/c^2$ 
and $120$~GeV$/c^2$.

\section{Results}
The systematic errors on the expected signal rate are:
(a) the luminosity measurement with less than 1~\%, (b)
the model dependence of the leptoquark fragmentation with
1 to 5~\%, (c) the parameter dependence for the jet finding with 2 to 5~\%, 
(d) the Monte Carlo statistics with 1~\%.

A special version of PYTHIA has been used to study the difference
between models where the hadronisation is simulated before and after
the leptoquark decay into an electron and a quark~\cite{bib-pylq}. 
The difference
in the average charged multiplicity within the geometrical
acceptance of the detector increases with
leptoquark mass and is always less than one unit. 
The model dependence of the leptoquark fragmentation has therefore been
estimated by varying the cut on the charged multiplicity by one unit in the
Monte Carlo while keeping it fixed in the data since the charged
multiplicity cut is expected to be very sensitive to the hadronisation
model. 
In addition,
the jet finding parameters have been varied: 
the value of the minimum transverse 
energy for a jet has been changed by $\pm$5~GeV and 
the jet radius $R$ has been changed from 1 to 0.7. The 
effect from the variation of $R$ is negligible.

The Monte Carlo model dependence has been studied by comparing
the alternative background Monte Carlo sets defined in 
section~\ref{sec-mc}. In the eq channel the alternative two-photon
and four-fermion sample predict each an increase in the
total number of background events of about $10\%$. This
would lead to a higher limit on the leptoquark mass 
than the Monte Carlo sample used.
For the $\nu$q-channel the four-fermion generators yield
consistent results and the expected two-photon background decreases
from 2.4 to 1.3 events, leading to a negligible change in the mass limits.

The systematic errors are added in quadrature and they
are taken into account in the limit
using the procedure of Highland and Cousins~\cite{bib-syst}.

The limits have been calculated for three different values of
the branching ratio $\beta=1$, $\beta=0.5$, and $\beta\to 0$, since 
for $\beta\equiv 0$ no production in eq collisions is possible.
The limit calculations have been performed according to the
procedure of~\cite{bib-bock} which takes into account the 
expected background, the mass resolution, the candidates,
and the efficiencies. 
The cross-section excluded at the 95\% confidence level, $\sigma_{95}$,
resulting from these calculations is shown in
Fig.~\ref{fig-cl} for scalar and for vector leptoquarks.
The resulting mass limits are given in Table~\ref{tab-limscalar} 
for $\lambda=\sqrt{4\pi\alpha_{\rm em}}$, where the electromagnetic coupling 
constant $\alpha_{\rm em}$ is taken at the
mass of the leptoquark with $\alpha_{\rm em}(M_{\rm LQ})\approx 1/128$. 
The upper limit at the $95~\%$ CL of the coupling
$\lambda$ ($\lambda'$) as a function of the mass $M_{\rm LQ}$ is
given in Figs.~\ref{13scalar} and \ref{53scalar}.
\renewcommand{\arraystretch}{1.0}
\begin{table}[htbp]
  \begin{center}
    \begin{tabular}{|c|c|c|c|c|c|}    
  \hline
 charge & state & $\beta=1$ & $\beta=0.5$ & $\beta\rightarrow 0$ 
\\ \hline
\rule[0.5cm]{0cm}{0cm} 
$\pm 1/3$ & S$_{\rm 0}$, S$_{\rm 1}$, \~S$_{1/2}$,  
& 163 GeV$/c^2$ & 158 GeV$/c^2$ & 175 GeV$/c^2$ \\ \hline
\rule[0.5cm]{0cm}{0cm} $-5/3$ & S$_{\rm 1/2}$ & 164 GeV$/c^2$ & - & -
\\ \hline
\rule[0.5cm]{0cm}{0cm} $-4/3$ & \~S$_{\rm 0}$ & 149 GeV$/c^2$ & - & -
\\ \hline
\rule[0.5cm]{0cm}{0cm} $-4/3$ & S$_{\rm 1}$ & 156 GeV$/c^2$ & - & -
\\ \hline
\rule[0.5cm]{0cm}{0cm} $-2/3$ & \~S$_{\rm 1/2}$,  
S$_{\rm 1/2}$, S$_{1/2}$ & 121 GeV$/c^2$ & - & 141 GeV$/c^2$ 
\\ \hline
\rule[0.8cm]{0cm}{0cm} $+2/3$ & S$_{1}$  & - & - & 162 GeV$/c^2$ 
\\ \hline
    \end{tabular}
  \caption{Mass limits for scalar leptoquarks and squarks for $\lambda=\sqrt{4\pi\alpha_{\rm em}}$ and the different $\beta$ values.}
    \label{tab-limscalar}
  \end{center}
\end{table}

\begin{figure}[htbp]
   \begin{center}
      \makebox[14cm][r]{
          \epsfxsize=14cm
          \epsffile{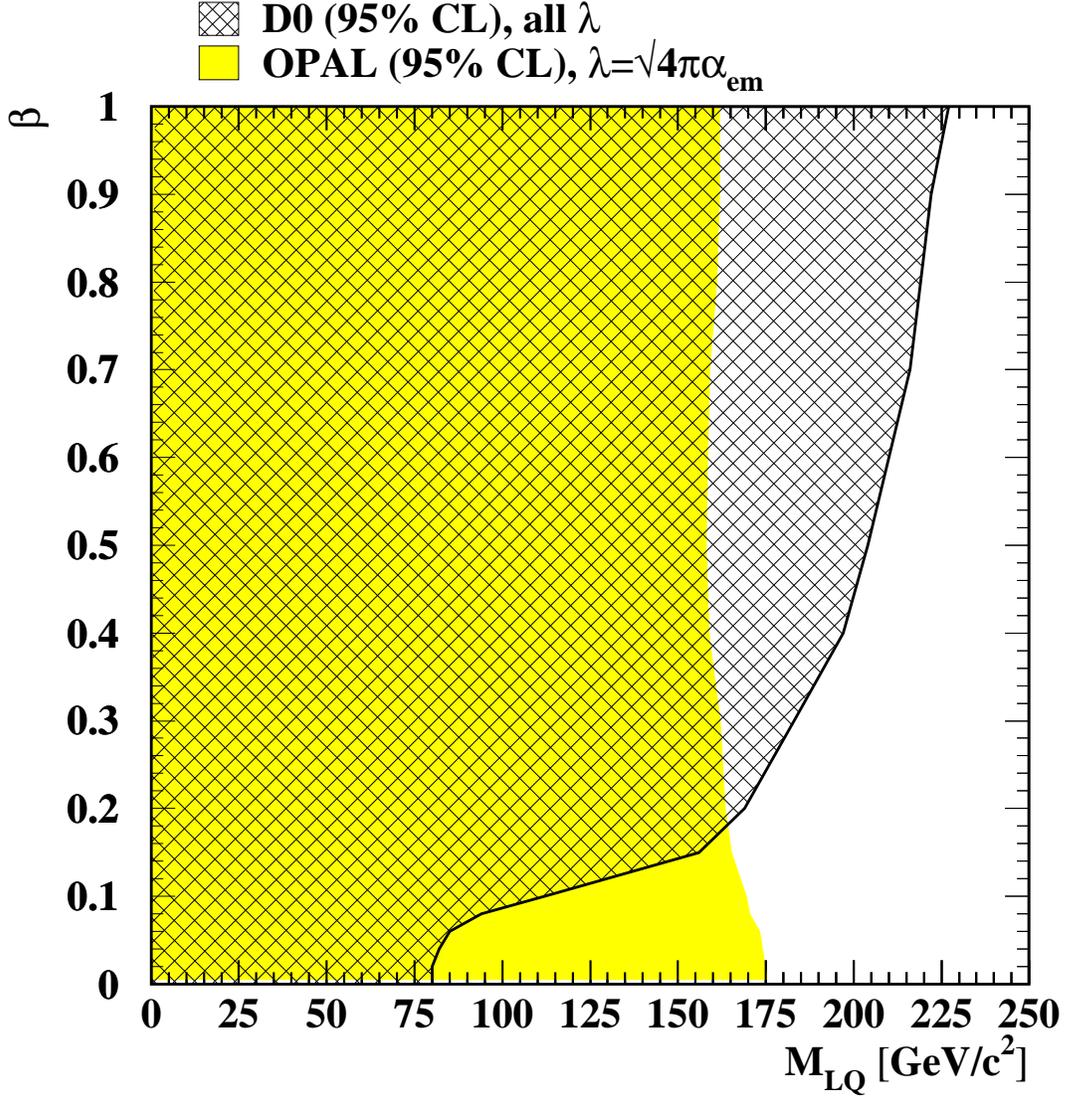}
          }
\caption{
Limit on $M_{\rm LQ}$ as a function of $\beta$
for couplings of electromagnetic strength 
($\lambda=\sqrt{4\pi\alpha_{\rm em}}$) for the charge $\pm 1/3$ scalar states.
The limit on $\beta$ is compared to the D0 results~\protect\cite{bib-d0}.
For $\beta\equiv 0$ no production in eq collisions is possible. 
}
\label{fig-beta}
\end{center}
\end{figure}

We have also derived the limit on $M_{\rm LQ}$ as a function of $\beta$
for couplings of electromagnetic strength 
($\lambda=\sqrt{4\pi\alpha_{\rm em}}$), i.e.~the assumption is
dropped that the branching ratio $\beta$ can take only the values $0,0.5$ and
$1$.
In Fig.~\ref{fig-beta} the limit on $\beta$ as a function
of $M_{\rm LQ}$ is compared to the D0 
results~\cite{bib-d0} which only exclude values up to 
$M_{\rm LQ}=80$~GeV$/c^2$ for $\beta=0$. Our analysis
is more sensitive in the low $\beta$ region, yielding
a limit of $M_{\rm LQ}=175$~GeV$/c^2$ for $\beta\to 0$. A similar
region has recently been explored by H1~\cite{bib-h1lq}.

Exactly 
the same procedure as for scalar leptoquarks has been used to determine 
the mass limits and limits on the Yukawa coupling for vector leptoquarks. 
The systematic errors are also identical. 
The results for the mass limits are shown in Table~\ref{tab-limvector}. 
The upper limit at the $95~\%$ CL of the coupling
$\lambda$ as a function of the mass $M_{\rm LQ}$ is
given in Figs.~\ref{v13}-\ref{v53}.

\renewcommand{\arraystretch}{1.0}
\begin{table}[htbp]
  \begin{center}
    \begin{tabular}{|c|c|c|c|c|}    
  \hline
 charge & state & $\beta=1$ & $\beta=0.5$ & $\beta\rightarrow 0$ \\ \hline
\rule[0.5cm]{0cm}{0cm} $-1/3$ & V$_{1/2}$, \~V$_{\rm 1/2}$ & 176 GeV$/c^2$ & 
- & 182 GeV$/c^2$ \\ \hline
\rule[0.5cm]{0cm}{0cm} $+1/3$ & V$_1$ & - & - & 188 GeV$/c^2$ \\ \hline
\rule[0.5cm]{0cm}{0cm} $-5/3$ & \~V$_0$ & 177 GeV$/c^2$ & - & -\\ \hline
\rule[0.5cm]{0cm}{0cm} $-5/3$ & V$_1$ & 182 GeV$/c^2$ & - & -\\ \hline
\rule[0.5cm]{0cm}{0cm} $-4/3$ & V$_{\rm 1/2}$ & 152 GeV$/c^2$ & - & -\\ \hline
\rule[0.5cm]{0cm}{0cm} $\pm 2/3$ & V$_0$, V$_1$, \~V$_{1/2}$ & 151 GeV$/c^2$ & 149 GeV$/c^2$ & 163 GeV$/c^2$ \\ \hline
    \end{tabular}
  \caption{Mass limits for vector leptoquarks for $\lambda=\sqrt{4\pi\alpha_{\rm em}}$ and the different possible $\beta$ values.}
    \label{tab-limvector}
  \end{center}
\end{table}

\subsection{Conclusions}
\label{sec-conclusions}
We have searched for singly-produced leptoquarks in 
electron-photon interactions at an e$^+$e$^-$ centre-of-mass energy 
of 189 GeV using the data collected with the OPAL detector at LEP. 
The data correspond to an integrated luminosity of about 160~pb$^{-1}$.
Some scalar leptoquark states can also be identified with
squarks in R-parity violating SUSY models. 
No evidence was found for the production of these particles. 
Therefore, $95\%$ confidence
limits were set on the mass as well as on the Yukawa coupling $\lambda$ for 
scalar and vector leptoquarks
and $\lambda'$ for squarks as a function of the
mass for different branching fractions $\beta$ in eq final states.

\section*{Acknowledgements}
We thank M.~Doncheski, S.~Godfrey and U.~Katz for
very useful discussions and 
C.~Papadopoulos for his help in using ERATO-LQ.\\
We particularly wish to thank the SL Division for the efficient operation
of the LEP accelerator at all energies
 and for their continuing close cooperation with
our experimental group.  We thank our colleagues from CEA, DAPNIA/SPP,
CE-Saclay for their efforts over the years on the time-of-flight and trigger
systems which we continue to use.  In addition to the support staff at our own
institutions we are pleased to acknowledge the  \\
Department of Energy, USA, \\
National Science Foundation, USA, \\
Particle Physics and Astronomy Research Council, UK, \\
Natural Sciences and Engineering Research Council, Canada, \\
Israel Science Foundation, administered by the Israel
Academy of Science and Humanities, \\
Minerva Gesellschaft, \\
Benoziyo Center for High Energy Physics,\\
Japanese Ministry of Education, Science and Culture (the
Monbusho) and a grant under the Monbusho International
Science Research Program,\\
Japanese Society for the Promotion of Science (JSPS),\\
German Israeli Bi-national Science Foundation (GIF), \\
Bundesministerium f\"ur Bildung und Forschung, Germany, \\
National Research Council of Canada, \\
Research Corporation, USA,\\
Hungarian Foundation for Scientific Research, OTKA T-029328, 
T023793 and OTKA F-023259.\\

\newpage

\begin{figure}[htbp]
   \begin{center}
          \epsfxsize=13.5cm
          \epsffile{pr337_06.epsi}
   \end{center}
\caption{({\bf a}) Selection efficiency in the $\nu$q channel 
after all cuts for scalar (dots) and 
vector (squares) leptoquarks;
({\bf b}) transverse mass $M_{\rm T}=2\ETMISS$ after all cuts 
for data (points with error bars), Standard Model background (full line) and 
two different scalar leptoquark 
masses (dotted line for 80 GeV$/c^2$ and dashed line for 120 GeV$/c^2$). 
The state S$_{1/2}$ (charge -2/3) was chosen to normalise the signal, 
using $\lambda=\sqrt{4\pi\alpha_{\rm em}}$.}
\label{fig-mnuqscalar}
%
   \begin{center}
          \epsfxsize=13.5cm
          \epsffile{pr337_07.epsi}
\caption{Cross-section excluded at the 95\% confidence level,
$\sigma_{95}$, using 
the number of candidates in the data for each channel separately, 
corresponding to $\beta$=1 for the electron-quark channel (full line) 
and $\beta$=0 for the neutrino-quark channel (dashed line), 
as well as for equal branching ratio into both channels 
($\beta$=0.5, dotted line). 
The expected SM background, the mass resolution, the candidates,
and the efficiencies are taken into account in the calculation.
The cross-section for the production of the
states a) S$_{1/2}$ (charge $-5/3$) and b) \~V$_{0}$ are also shown.}
\label{fig-cl}
\end{center}
\end{figure}

\begin{figure}[hbtp]
   \begin{center}
          \epsfxsize=14cm
          \epsffile{pr337_08.epsi}
\caption{Limits on the coupling constants $\lambda$ ($\lambda_{1jk}'$) 
for scalar leptoquark states (squarks) with 
({\bf a}) charge $\pm\frac{1}{3}$: S$_0$ with $\beta=1$ (full line),  
S$_{\rm 0}$, ${\rm \tilde{d}_R}$, S$_{\rm 1}$  with $\beta=0.5$ (dashed line) 
as well as \~S$_{1/2}$ and ${\rm \bar{\rm \tilde{d}}_L}$ with $\beta=0$ 
({\bf b}) charge $-\frac{5}{3}$: S$_{1/2}$ with $\beta=1$.}
\label{13scalar}
\vspace{0.3cm}
          \epsfxsize=14cm
          \epsffile{pr337_09.epsi}
\caption{Limits on the coupling constant $\lambda$ ($\lambda_{1jk}'$) for 
scalar leptoquark states (squarks) with 
({\bf a}) charge $-\frac{4}{3}$: S$_1$ with $\beta=1$ (full line),  
\~S$_{\rm 0}$,  
with $\beta=1$ (dashed line) ({\bf b}) charge $\pm\frac{2}{3}$: 
\~S$_{\rm 1/2}$, 
${\rm \bar{\rm \tilde{u}}_L}$, S$_{\rm 1/2}$ with $\beta=1$ (full line), 
S$_{1/2}$ with  $\beta=0$ (dotted line) and S$_1$ with  $\beta=0$ 
(dashed-dotted line).}
\label{53scalar}
\end{center}
\end{figure}

\begin{figure}[htbp]
   \begin{center}
          \epsfxsize=14cm
          \epsffile{pr337_10.epsi}
\caption{Limits on the coupling constant $\lambda$ for vector leptoquark states with 
({\bf a}) charge $\pm\frac{1}{3}$: V$_{1/2}$,\~V$_{1/2}$ with $\beta=1$ (full line) 
and V$_{1/2}$, V$_{1}$  with $\beta=0$ 
({\bf b}) charge $-\frac{5}{3}$: \~V$_{0}$ with $\beta=1$ (full line) and 
V$_1$ with $\beta=1$ (dashed line).}
\label{v13}
\vspace{0.3cm}
          \epsfxsize=14cm
          \epsffile{pr337_11.epsi}
\caption{Limits on the coupling constant $\lambda$ for vector leptoquark states with 
({\bf a}) charge $-\frac{4}{3}$: V$_{1/2}$ with $\beta=1$ (full line) 
({\bf b}) charge $\pm\frac{2}{3}$: V$_{0}$ with $\beta=1$ (full line), V$_{0}$, 
V$_{1}$ with  $\beta=0.5$ (dashed line) and V$_{1/2}$ with  $\beta=0$ 
(dotted line).}
\label{v53}
\end{center}
\end{figure}

\end{document}